\documentclass{PoS}

\usepackage{graphicx,xcolor}
\usepackage{epsf}
\usepackage{amsmath}
\usepackage{amssymb}
\usepackage{booktabs}
\usepackage{siunitx}
\usepackage{listings}
\newcommand{\beq}{\begin{equation}}
\newcommand{\eeq}{\end{equation}}
\newcommand{\bea}{\begin{eqnarray}}
\newcommand{\eea}{\end{eqnarray}}

\newcommand{\vsb}{\vspace{-0.1cm}}

\title{BFKL effects and central rapidity dependence 
       in Mueller-Navelet jet production at 13 TeV LHC}

\ShortTitle{BFKL effects and central rapidity dependence 
            in Mueller-Navelet jet production at 13 TeV LHC}

\author{\speaker{Francesco Giovanni Celiberto} \\
        Dipartimento di Fisica, Università della Calabria
        and Istituto Nazionale di Fisica Nucleare, 
        Gruppo Collegato di Cosenza,
        Arcavacata di Rende, 87036 Cosenza, Italy \\
        E-mail: 
        \email{francescogiovanni.celiberto@fis.unical.it}}

\author{Dmitry Yu. Ivanov \\
        Sobolev Institute of Mathematics, 
        630090 Novosibirsk, Russia \\
        Novosibirsk State University, 
        630090 Novosibirsk, Russia \\
        E-mail: 
        \email{d-ivanov@math.nsc.ru
}}
        
\author{Beatrice Murdaca \\
        Istituto Nazionale di Fisica Nucleare, \\
        Gruppo Collegato di Cosenza, 
        Arcavacata di Rende, 87036 Cosenza, Italy \\
        E-mail: 
        \email{beatrice.murdaca@cs.infn.it
}}
        
\author{Alessandro Papa \\
        Dipartimento di Fisica, Università della Calabria
        and Istituto Nazionale di Fisica Nucleare, 
        Gruppo Collegato di Cosenza,
        Arcavacata di Rende, 87036 Cosenza, Italy \\
        E-mail: 
        \email{alessandro.papa@fis.unical.it}}

\abstract{A study of the production of Mueller-Navelet jets 
          at 13 TeV LHC is presented, including 
          BFKL resummation effects and investigating 
          three different variants of the BLM scale 
          optimization method. It is shown how the 
          cross section and the azimuthal observables 
          are affected by the exclusion of the events 
          where, for a given rapidity interval between 
          the two jets, one of these is produced in the
          central region.}

\FullConference{XXIV International Workshop on 
                Deep-Inelastic Scattering 
                and Related Subjects\\
		11-15 April, 2016\\
		DESY Hamburg, Germany}

\begin{document}

\section{Introduction}
\label{intro}

The production at the LHC of Mueller-Navelet jets~\cite{Mueller:1986ey} 
represents a fundamental test of QCD at high energies.
It is an inclusive process where two jets, characterized by large transverse 
momenta that are of the same order and much larger than $\Lambda_{\rm QCD}$, 
are produced in proton-proton collisions, separated by a large rapidity gap 
$Y$ and in association with an undetected hadronic system $X$.
At the LHC energies the rapidity gap between the two jets can be large enough,
that the emission of several undetected hard partons, having large transverse 
momenta, with rapidities intermediate to those of the two detected jets, 
becomes possible. 

The BFKL approach~\cite{BFKL} provides with a systematic framework
for the resummation of the the energy logarithms 
that accompany this undetected parton radiation, 
both in the leading logarithmic approximation (LLA) 
and in the next-to-leading logarithmic approximation (NLA).
In this approach, the cross section for Mueller-Navelet jet production takes
the form of a convolution between two impact factors for the transition from
each colliding proton to the forward jet (the so-called ``jet vertices'') and
a process-independent Green's function~\cite{FL98,CC98,Fadin:1998jv,FG00,FF05}.
The jet vertex can be expressed, within collinear factorization at the
leading twist, as the convolution of parton distribution functions (PDFs)
of the colliding proton, obeying the standard DGLAP evolution~\cite{DGLAP},
with the cross section of hard process describing the transition from the parton emitted
by the proton to the forward jet in the final state. The Mueller-Navelet
jet production process is, therefore, a unique venue, where the two main
resummation mechanisms of perturbative QCD play their role at the same time.
The expression for the ``jet vertices'' was first obtained with NLO accuracy
in~\cite{Bartels:2002yj}, a result later confirmed in~\cite{Caporale:2011cc}.
A simpler expression, more practical for numerical purposes, was
obtained in~\cite{Ivanov:2012ms} within the so-called ``small-cone''
approximation (SCA)~\cite{Furman:1981kf,Aversa}.
A lot of papers have appeared, so far, 
about the Mueller-Navelet jet production
process at LHC, both at a center-of-mass energy of
14~TeV~\cite{Colferai2010,Caporale2013,Salas2013} and
7~TeV~\cite{Ducloue2013,Ducloue2014,Ducloue:2014koa,
            Caporale:2014gpa,Celiberto:2015yba}. 
Their main aim was the study of the $Y$-dependence of azimuthal angle
correlations between the two measured jets and of ratios between them~\cite{sabioV}.
In order to improve the perturbative stability of the BFKL series, 
several possibilities 
were considered, such as 
collinear improvement~\cite{collinear}, 
energy-momentum conservation~\cite{Kwiecinski:1999yx}, 
PMS~\cite{PMS}, FAC~\cite{FAC}, and BLM~\cite{BLM}.
There is a clear evidence that theoretical results can nicely reproduce CMS
data~\cite{CMS} at 7~TeV in the range $5 \lesssim Y \lesssim 9.4$ when the
BLM optimization method is adopted.

The large rapidity gaps provided by the LHC definitely offer us 
a unique opportunity to disentangle the applicability region of
the high-energy resummation. 
To this aim, new ways to probe BFKL have been recently 
investigated. On one side, one can study a process featuring 
a less inclusive final state, by allowing the detection of two 
charged light hadrons - instead of two jets - 
separated by a large interval 
of rapidity~\cite{Ivanov:2012iv,Celiberto:2016hae}.
On the other side, it was suggested to study 
azimuthal correlations where transverse momenta 
and azimuthal angles of extra particles
introduce a new dependence. Therefore, 
the study of three and four-jet production processes 
has been proposed as a novel possibility to define 
new, generalized and suitable BFKL 
observables~\cite{Caporale:2015vya,Caporale:2015int,
Caporale:2016soq,Caporale:2016xku,Celiberto:2016vhn}.

Returning back to Mueller-Navelet jets, 
there is another issue which has not been taken
into account both in theoretical and experimental analyses so far.
The rapidity of one of the two jets could be so small, say $|y_{J_i}|\lesssim 2$, 
that this jet is actually produced in the central region, 
rather than in one of the two forward regions. 
In this kinematic region PDF parametrizations 
extracted in NNLO and in NLO approximations start to differ 
one from the other. Recently, in Ref.~\cite{Currie:2013dwa}, results 
for NNLO corrections to the dijet production originated from the gluonic 
subprocesses were presented. In the region $|y_{J_{1,2}}|<0.3$ and for jet 
transverse momenta $\sim 100$~GeV, the account of NNLO effects leads to an
increase of the cross section by $\sim 25\%$. For our kinematics, 
featuring smaller jet transverse momenta 
and ``less inclusive'' coverage of jet 
rapidities, one could expect even larger NNLO corrections.
In view of this statement, we propose 
to return to the original Mueller-Navelet idea, 
to study the inclusive production 
of two forward jets separated by a large rapidity gap, 
and to remove from the 
analysis those regions where jets 
are produced at central rapidities by 
imposing the constraint that the rapidity 
of a Mueller-Navelet jet 
cannot be smaller than a given value. 


\section{Theoretical setup}
\label{theory}

We consider the production of Mueller-Navelet
jets~\cite{Mueller:1986ey} in proton-proton collisions
\begin{equation}
\label{process}
p(p_1) + p(p_2) \to {\rm jet}(k_{J_1}) + {\rm jet}(k_{J_2})+ X \;,
\end{equation}
where the two jets are characterized 
by high transverse momenta,
$\vec k_{J_1}^2\sim \vec k_{J_2}^2\gg \Lambda_{\rm QCD}^2$ 
and large separation
in rapidity, while $p_1$ and $p_2$ are taken 
as Sudakov vectors.

In QCD collinear factorization the cross section 
of the process~(\ref{process}) reads
\beq
\frac{d\sigma}{dx_{J_1}dx_{J_2}d^2k_{J_1}d^2k_{J_2}}
=\sum_{i,j=q,{\bar q},g}\int_0^1 dx_1 
\int_0^1 dx_2\ f_i\left(x_1,\mu_F\right)
\ f_j\left(x_2,\mu_F\right)
\frac{d{\hat\sigma}_{i,j}\left(x_1x_2s,\mu_F\right)}
{dx_{J_1}dx_{J_2}d^2k_{J_1}d^2k_{J_2}}\;,
\eeq
where the $i, j$ indices specify 
the parton types (quarks $q = u, d, s, c, b$; antiquarks 
$\bar q = \bar u, \bar d, \bar s, \bar c, \bar b$; 
or gluon $g$), $f_{i,j}\left(x, \mu_F \right)$ 
denotes the initial proton PDFs; $x_{1,2}$ are
the longitudinal fractions of the partons involved 
in the hard subprocess,
while $x_{J_{1,2}}$ are the jet momenta longitudinal fractions;
$\mu_F$ is the factorization scale; 
$d\hat\sigma_{i,j}\left(x_1x_2s, \mu_F \right)$ is
the partonic cross section for the production of jets and
$x_1x_2s\equiv\hat s$ is the squared center-of-mass 
energy of the parton-parton collision subprocess 
(see Fig.~1 of Ref.~\cite{Celiberto:2016ygs}).
The cross section of the process can be written as
\beq
\frac{d\sigma}
{dy_{J_1}dy_{J_2}\, d|\vec k_{J_1}| \, d|\vec k_{J_2}|
d\phi_{J_1} d\phi_{J_2}}
=\frac{1}{(2\pi)^2}\left[{\cal C}_0+\sum_{n=1}^\infty  
2\cos (n\phi )\, {\cal C}_n\right]\, ,
\eeq
where $\phi=\phi_{J_1}-\phi_{J_2}-\pi$, 
while ${\cal C}_0$ gives the total
cross section and the other coefficients 
${\cal C}_n$ determine the distribution
of the azimuthal angle of the two jets.
To fix (at a common value) 
the renormalization and factorization scales, 
$\mu_R$ and $\mu_F$, 
we will make use of the ``exact'' and in some cases
also of two approximate, 
semianalytic implementations of BLM method,
labeled as $(a)$ and $(b)$, 
using the so called \emph{exponentiated} representation 
to keep contact with 
previous works~\cite{Caporale:2015uva}. 

\begin{figure}[t]
\centering
   \hspace{-0.8cm}
   \includegraphics[scale=0.235]
    {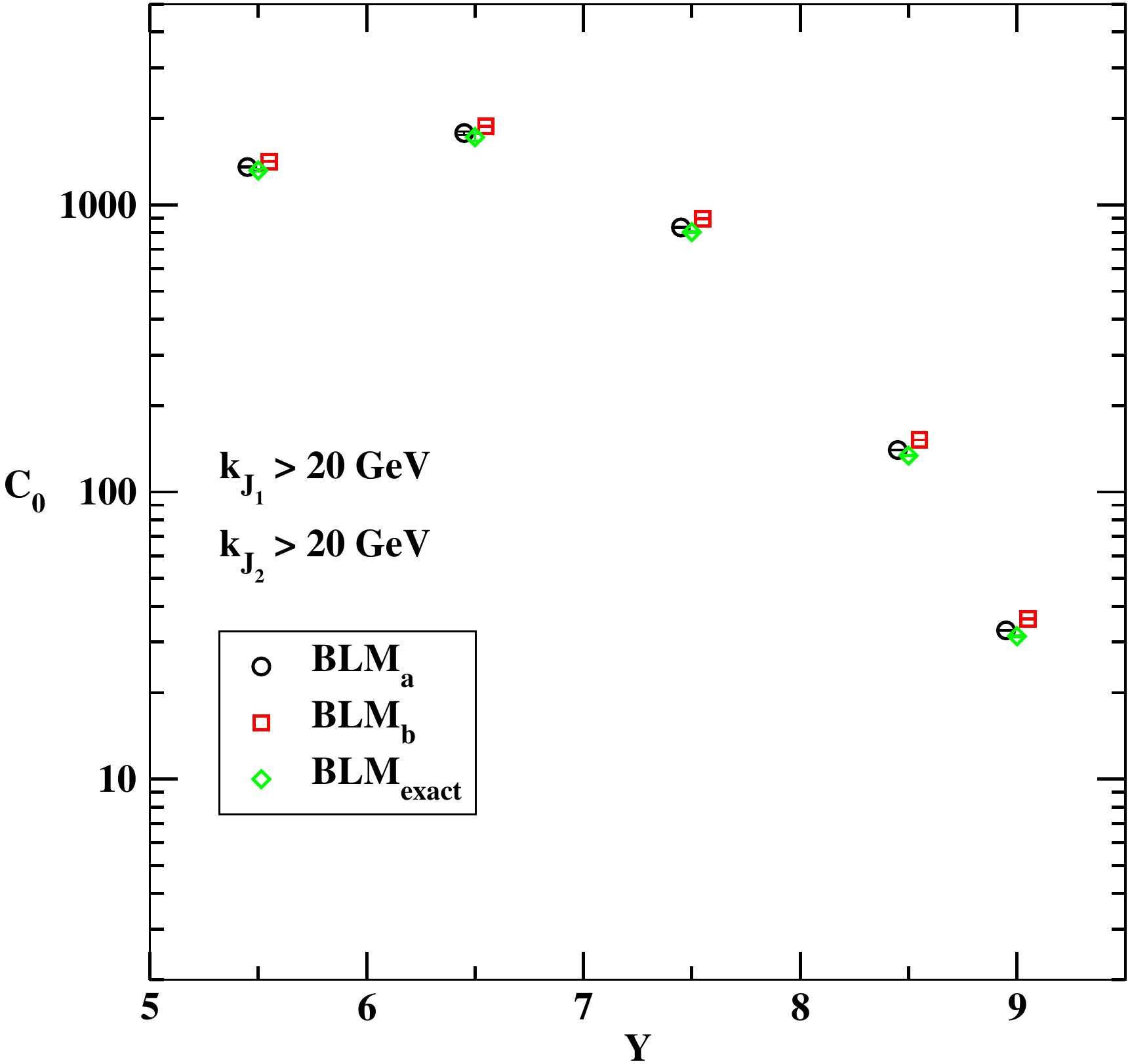}
   \hspace{0.5cm}
   \includegraphics[scale=0.235]
    {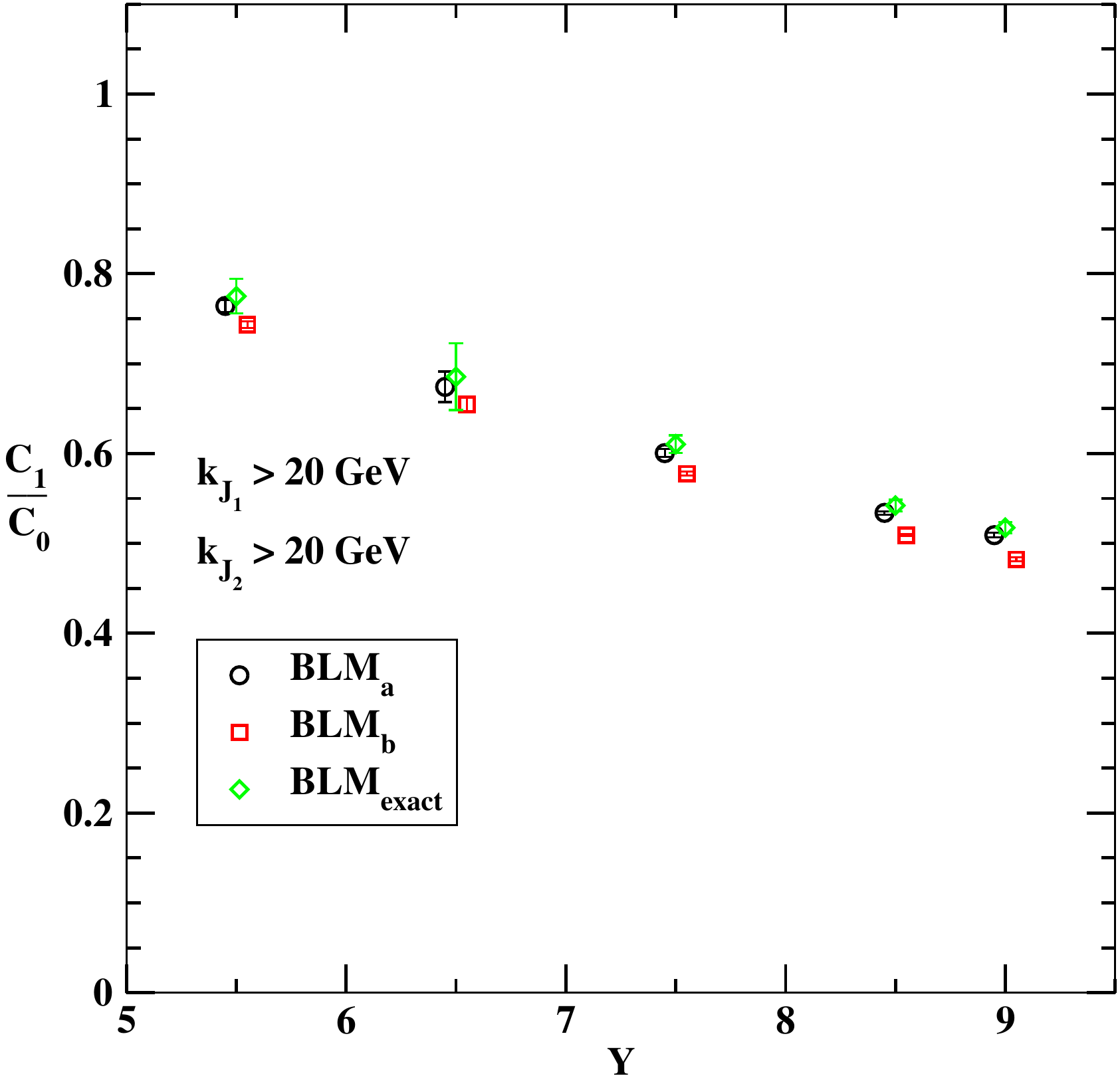}
   \hspace{0.5cm}
   \includegraphics[scale=0.235]
    {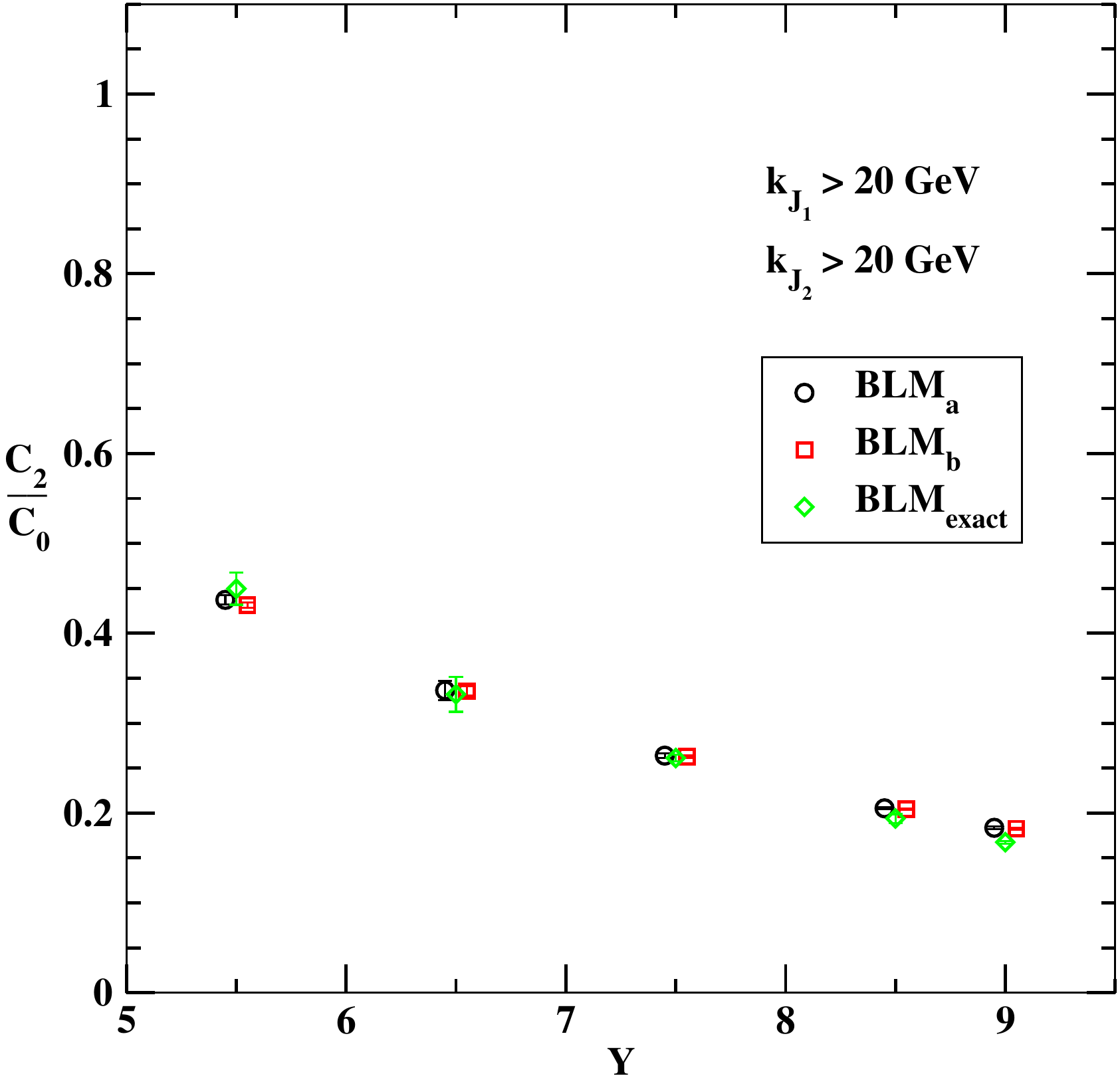}

   \includegraphics[scale=0.235]
    {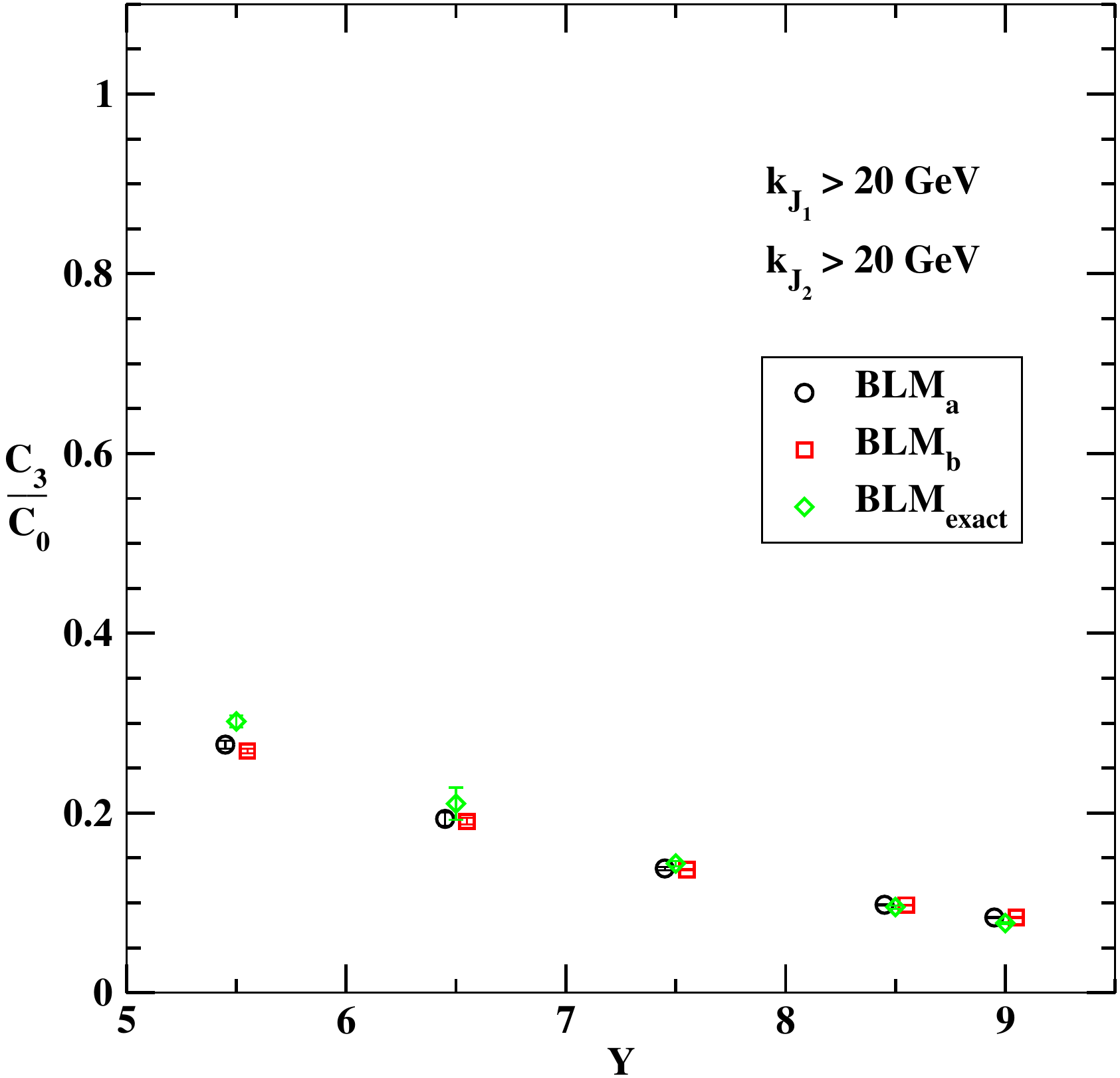}
   \hspace{0.5cm}
   \includegraphics[scale=0.235]
    {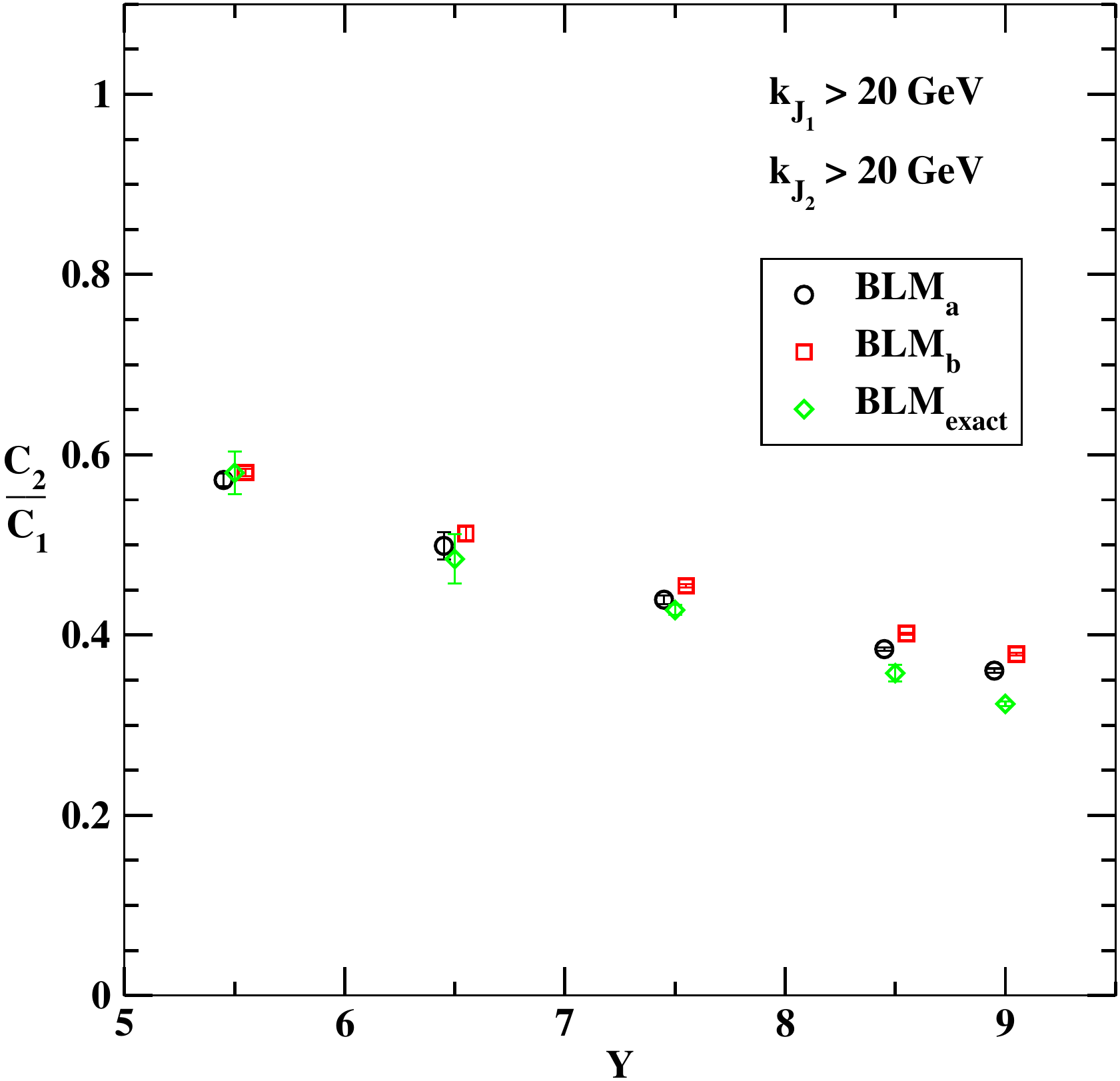}
   \hspace{0.5cm}
   \includegraphics[scale=0.24]
    {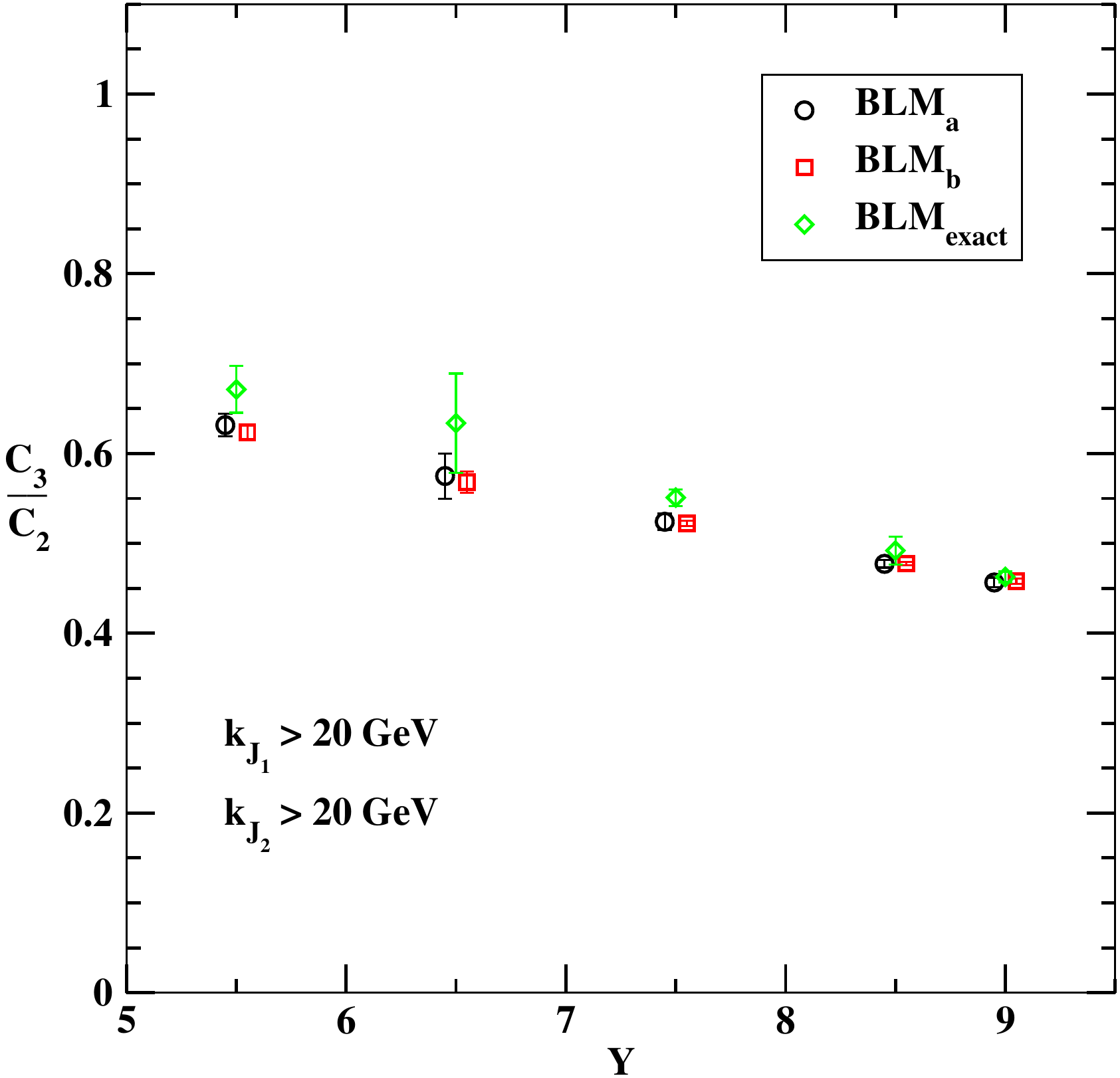}
   \hspace{0.5cm}
\caption{$Y$-dependence of $C_0$ 
and of several $R_{nm}$ ratios
for $k_{J_1,\rm min}=k_{J_2,\rm min}=20$~GeV 
and for $|y_{J_1}| > 2.5$, from
the three variants of the BLM method.
For the numerical values, 
see Table 1 of Ref.~\cite{Celiberto:2016ygs}.}
\label{2020_blm_abe}
\end{figure}

\begin{figure}[t]
\centering
   \hspace{-0.8cm}
   \includegraphics[scale=0.235]
    {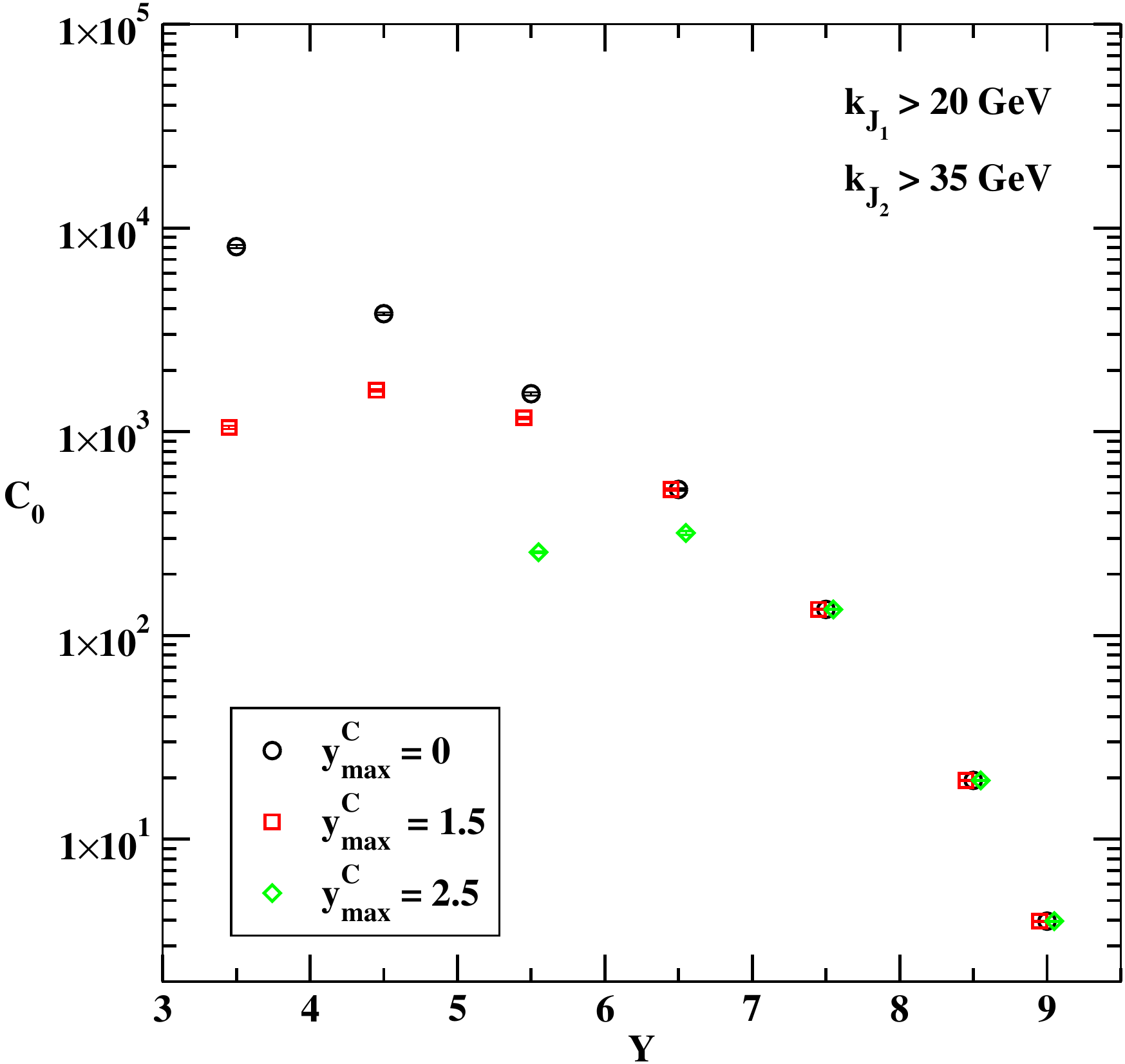}
   \hspace{0.5cm}
   \includegraphics[scale=0.235]
    {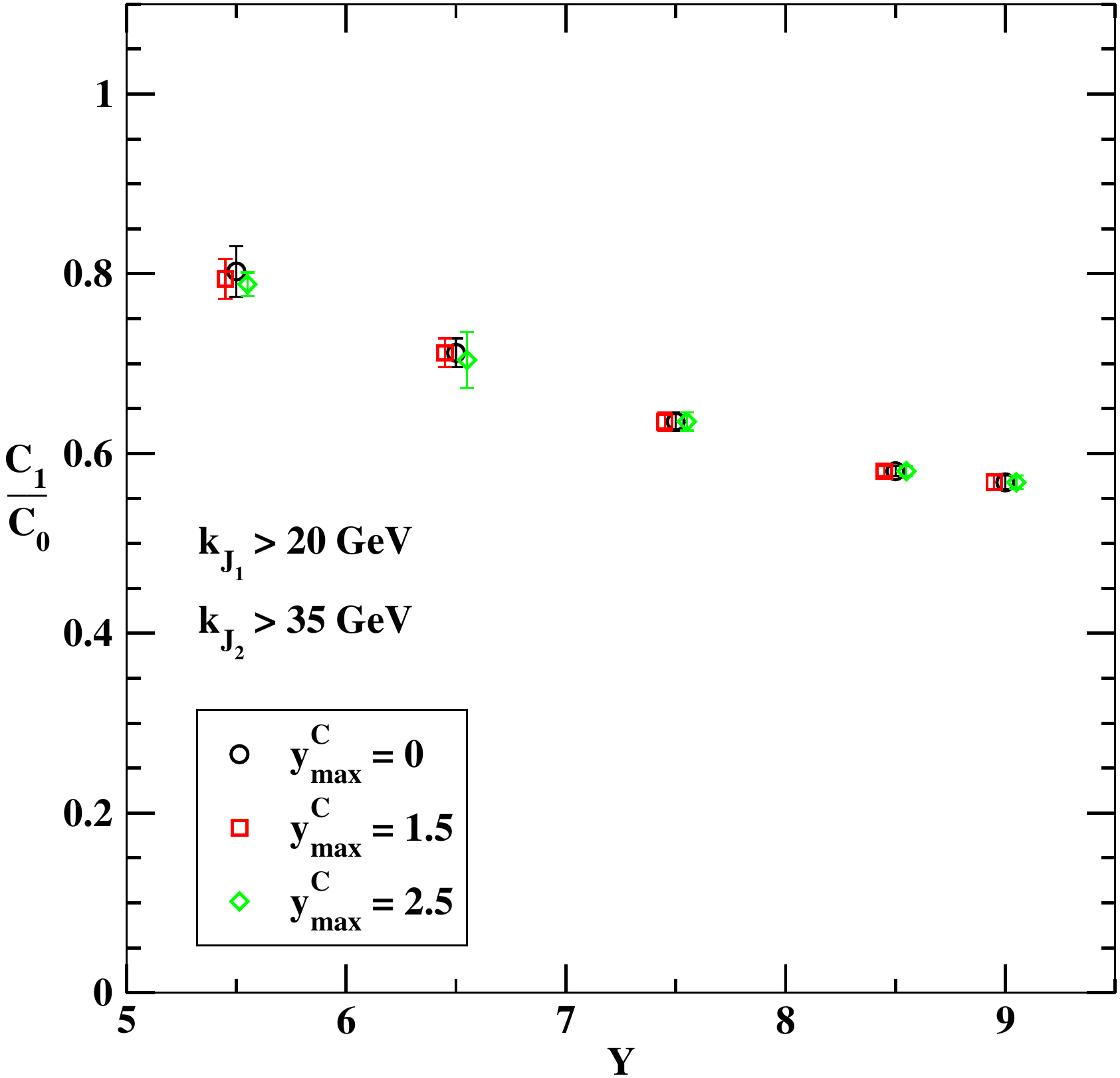}
   \hspace{0.5cm}
   \includegraphics[scale=0.235]
    {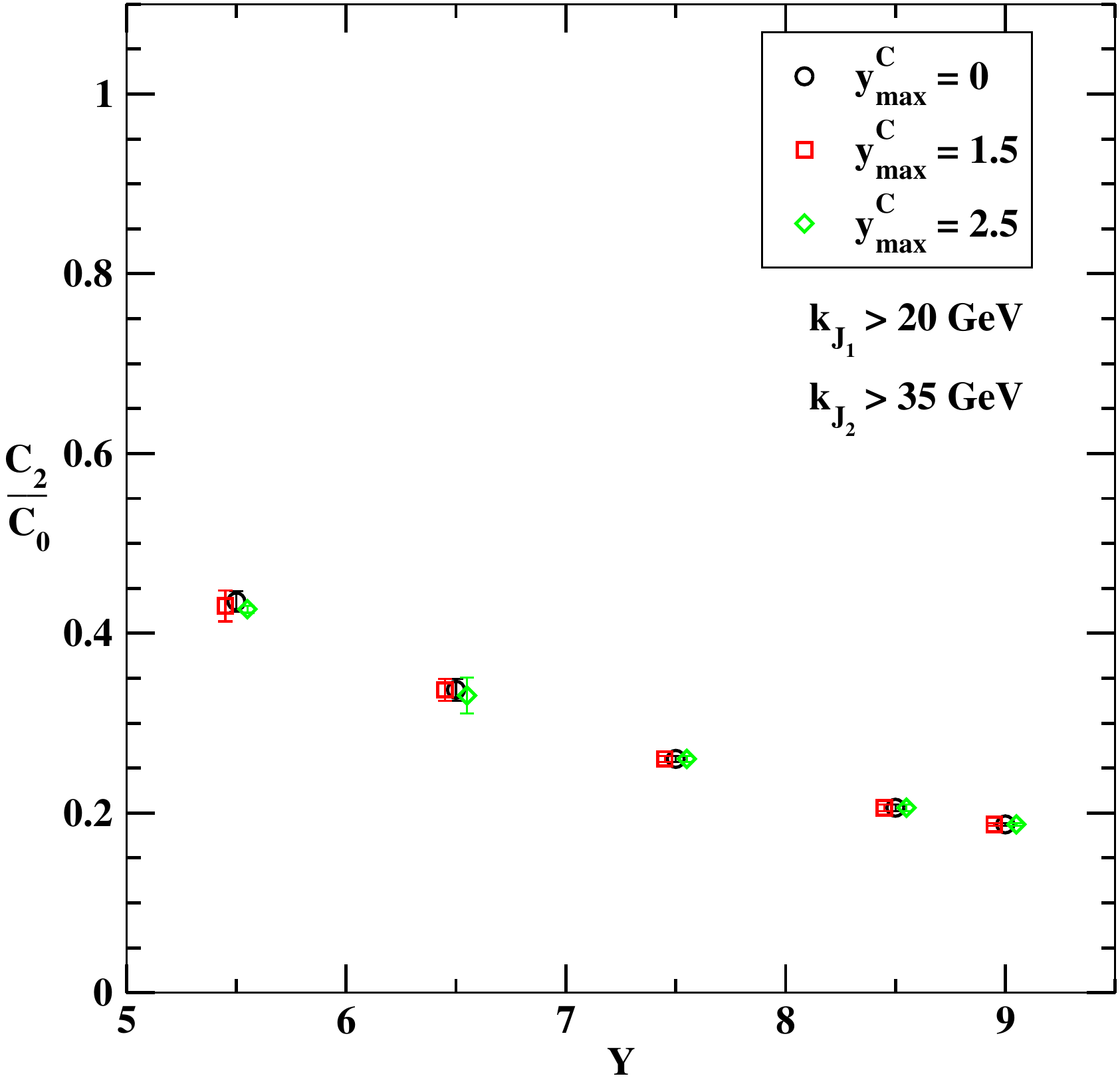}
   
   \includegraphics[scale=0.235]
    {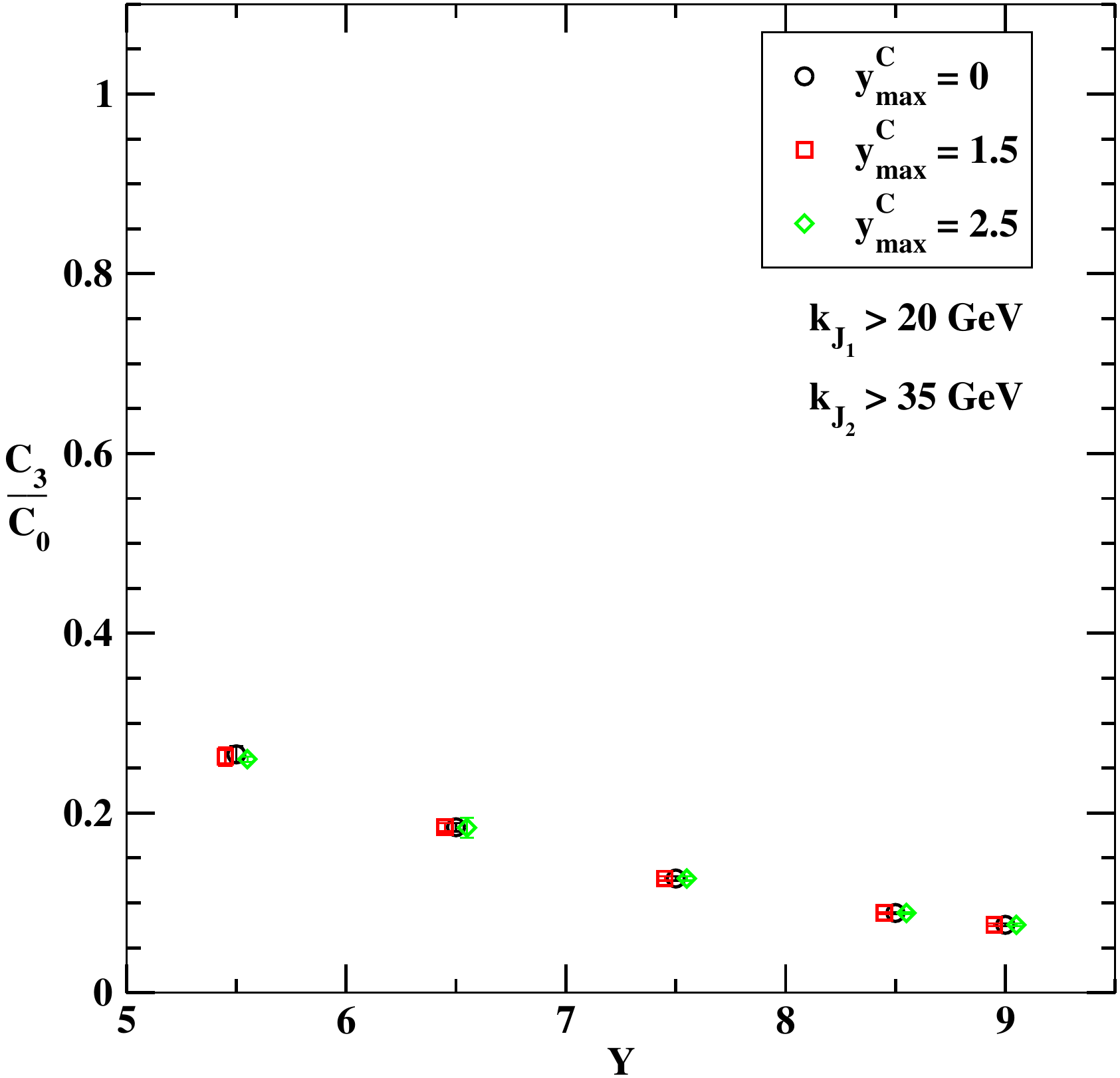}
   \hspace{0.5cm}    
   \includegraphics[scale=0.235]
    {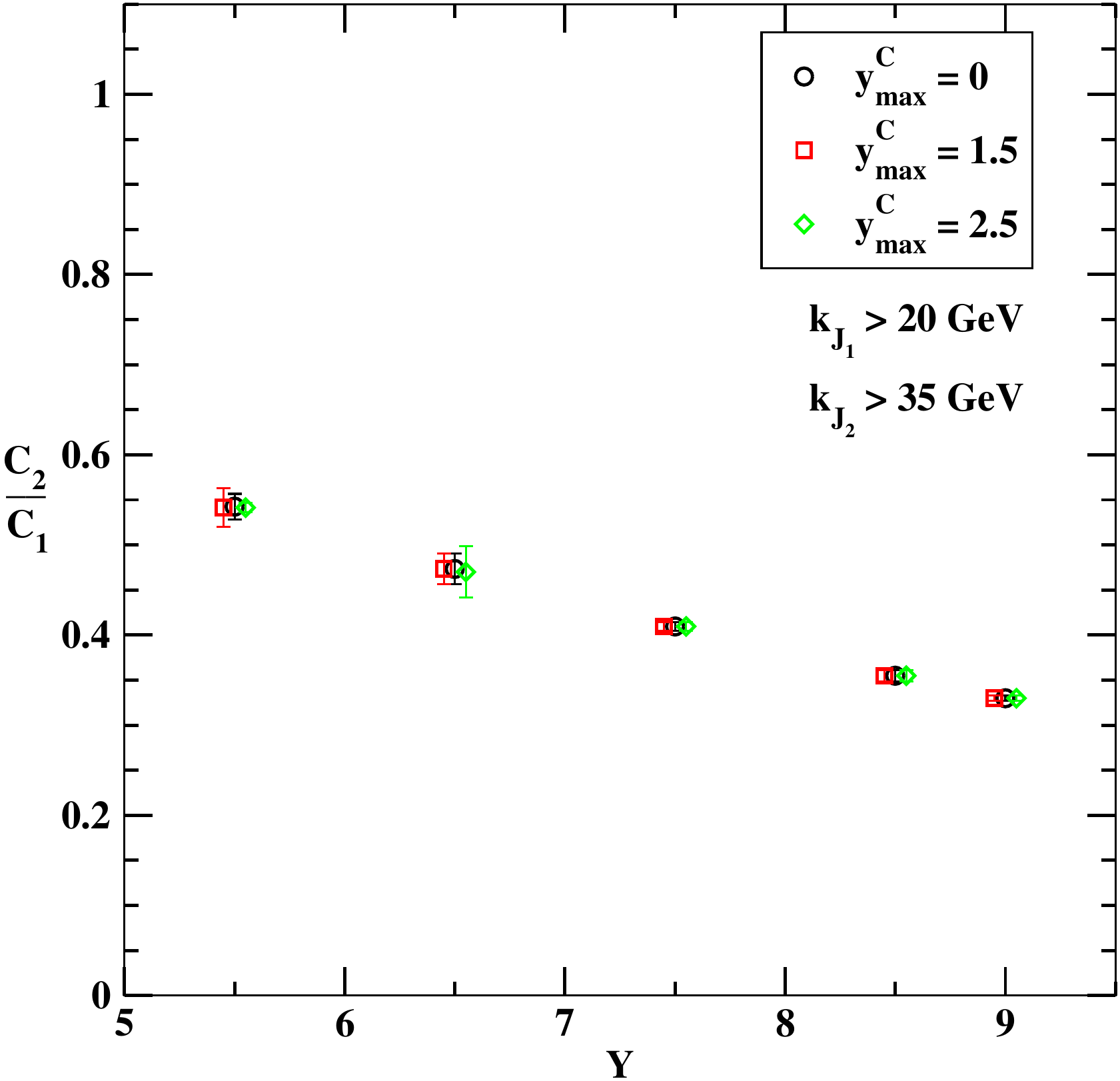}
   \hspace{0.5cm}
   \includegraphics[scale=0.235]
    {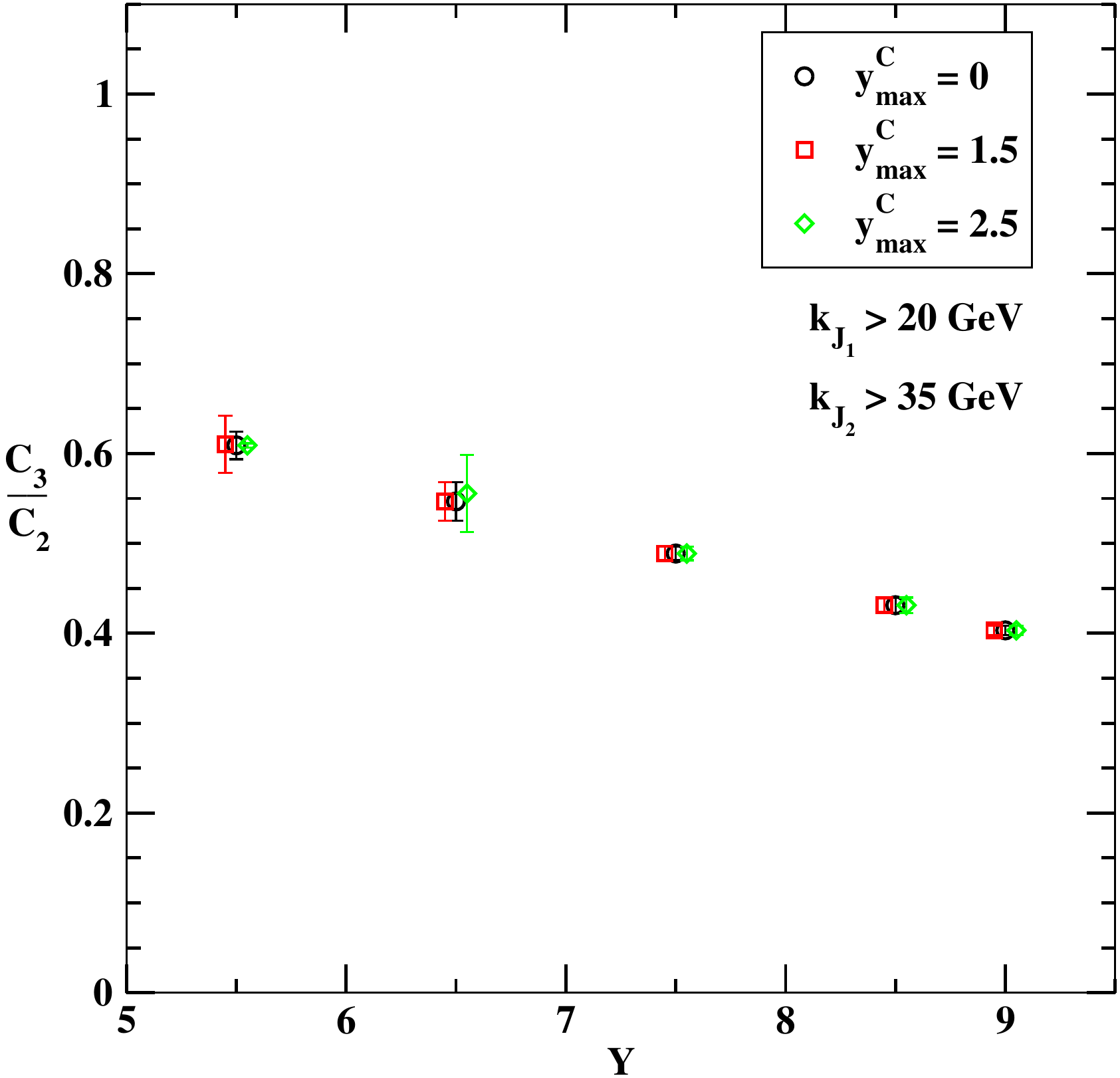}
   \hspace{0.5cm}
\caption{$Y$-dependence of $C_0$ 
and of several $R_{nm}$ ratios
from the ``exact'' BLM method, 
for $k_{J_1,\rm min}=20$~GeV, $k_{J_2,\rm min}=35$~GeV
and for all choices of the cuts 
of the central rapidity region.
For the numerical values, 
see Tables 2-7 of Ref.~\cite{Celiberto:2016ygs}.}
\label{2035_blm_exact}
\end{figure}


\section{Numerical analysis}
\label{results}

We present here our results for the dependence on the
rapidity separation between the detected jets, $Y=y_{J_1}-y_{J_2}$,
of ratios ${\cal R}_{nm}\equiv{\cal C}_n/{\cal C}_m$ between the
coefficients ${\cal C}_n$. Ratios of
the form ${\cal R}_{n0}$ have a simple physical interpretation, 
being the azimuthal correlations $\langle \cos(n\phi)\rangle$. 
To match the CMS kinematic cuts, we 
consider the \emph{integrated coefficients} given by
\begin{equation}\label{Cm_int}
C_n=
\int_{y_{1,\rm min}}^{y_{1,\rm max}}\hspace{-0.21cm}dy_1
\int_{y_{2,\rm min}}^{y_{2,\rm max}}\hspace{-0.21cm}dy_2
\int_{k_{J_1,\rm min}}^{\infty}\hspace{-0.21cm}dk_{J_1}
\int_{k_{J_2,\rm min}}^{\infty}\hspace{-0.21cm}dk_{J_2}
\delta\left(y_1-y_2-Y\right)
\theta\left(|y_1| - y^{\rm C}_{\rm max}\right) \theta\left(|y_2| - y^{\rm C}_{\rm max}
\right) {\cal C}_n 
\end{equation}
and their ratios $R_{nm}\equiv C_n/C_m$. In Eq.~(\ref{Cm_int}),
the two step-functions force the exclusion of jets with rapidity 
smaller than a cutoff value $y^{\rm C}_{\rm max}$, which delimits the central
rapidity region. We will take jet rapidities in the range delimited by
$y_{1,\rm min}=y_{2,\rm min}=-4.7$  and $y_{1,\rm max}=y_{2,\rm max}=4.7$, 
as in the CMS analyses at 7~TeV.
As for the central rapidity exclusion, 
we will consider the three cases 
$y^{\rm C}_{\rm max}=0$, 1.5, 2.5.
As for jet transverse momenta, differently from most previous analyses,
we make five different choices which include \emph{asymmetric} cuts.
The center-of-mass energy is fixed at $\sqrt s=13$ TeV.
For details on kinematic setup, 
numerical tools used and uncertainties,
see Ref.~\cite{Celiberto:2016ygs}.

The main result we have found is that, 
except for the case of total cross section, ${\cal C}_0$,  
$R_{nm}$ remain unaffected by the cut on the central
rapidity region, over the entire region of values of $Y$. 
This is obvious for the values of $Y$ large enough 
to be insensitive to the very presence 
of a non-zero $y^{\rm C}_{\rm max}$, 
but it is unexpectedly true also 
for the lower values of $Y$. 


\section{Summary}
\label{summary}

We have considered the Mueller-Navelet jet production
process at LHC at the center-of-mass energy of 13~TeV and have given
predictions for total cross sections 
and several moments of the jet azimuthal angle distribution,  
using the BLM method to optimize the $\mu_R$ and $\mu_F$ scales. 
Differently from previous studies, we have 
considered the effect of excluding that one 
of the two detected jets be produced 
in the central rapidity region.
Indeed, central jets originate 
from small-$x$ partons and the collinear approach for 
the description of the Mueller-Navelet jet vertices 
may be not good at small $x$. 
It would be very interesting 
to confront our predictions with LHC data. 

\vspace{-0.3cm}

\acknowledgments{
The work of D.I. was supported in part 
by the grant RFBR-15-02-05868-a.}

\end{document}